\documentstyle[12pt]{article}

\font\mybb=msbm10 at 11pt
\def\bb#1{\hbox{\mybb#1}}
\def\ZZ {\bb{Z}}

\def\exp{{\rm exp}}

\newcommand{\sectiono}[1]{\section{#1}\setcounter{equation}{0}}

\begin{document}
\thispagestyle{empty}
{\hfill CALT-68-1999

\hfill hep-th/9506076

\vspace {1.0cm}}
\parskip=9pt     % spacing between paragraphs
\begin{center}{\Large \bf Classical Symmetries of Some\break Two-Dimensional
Models Coupled to Gravity\footnote{Work supported in part
by U.S. Department of Energy
Grant No. DE-FG03-92-ER40701.}}
\end{center}
\bigskip
\bigskip
\centerline{\large John H. Schwarz\footnote{email: jhs@theory.caltech.edu}}
\medskip
\centerline{California Institute of Technology}
\centerline{Pasadena, CA 91125}
\bigskip
\parindent=1 cm

\begin{abstract}
This paper is a sequel to one in which we examined the affine symmetry
algebras of arbitrary  classical
principal chiral models and symmetric space models
in two dimensions. It examines the extension of those
results in the presence of gravity. The main result is that
even though the symmetry transformations
of the fields depend on the gravitational
background, the symmetry algebras of these classical theories
are completely unchanged by the presence of arbitrary
gravitational backgrounds. On the other hand, we are unable to generalize
the Virasoro symmetries of the flat-space theories to theories with gravity.
\end{abstract}
\vfil \eject

\sectiono{Introduction}

String theories possess large discrete symmetry  groups -- called dualities.
The first class of these to be understood were
``T dualities'' \cite{giveon} -- symmetries
associated with compactification, which are realized order-by-order in string
perturbation theory.  Subsequently, a class of symmetries called ``S
dualities,'' \cite{font}
a string generalization of the electric-magnetic duality of
certain field theories \cite{montonen},
was conjectured.  While they look quite similar to T
duality from the viewpoint of low-energy effective field theory, they are much
more speculative than T dualities.  The reason is that they are
non-perturbative symmetries of theories that are only known perturbatively!
While this means that we are really not yet in a position to prove that they
are true (though supporting ``evidence'' has been obtained \cite{sena}),
it also makes them
extremely interesting as a window into the deeper non-perturbative workings of
string theories.  In certain cases, such as toroidal compactification of type
II strings to four dimensions or the heterotic string to three dimensions, the
S
and T dualities are subgroups of a larger group of symmetries, known as
U dualities \cite{hull,senb}.
This is another indication that S and T duality are not really so
different.  Also, it has been conjectured that certain string theories have
dual formulations in which the roles of S and T duality are interchanged -- so
that S becomes perturbative and  T becomes nonperturbative.  This phenomenon is
called ``duality of dualities'' \cite{jhssen}.

A remarkable possibility suggested by recent works is that there is just one
superstring theory, and that the type I, type II, and heterotic theories,
compactified in various ways, are interrelated by a fascinating web of
non-perturbative transformations \cite{hull,witten}.
It is even conceivable that all
solutions could be part of a smoothly connected
moduli space \cite{strominger}.  This suggests
that someday it may be possible to identify a unique quantum ground state of a
unique theory!  This has been my dream for more than 20 years, though
developments during the period 1985-88 made it appear very unlikely.  Today the
outlook is much brighter.

The viewpoint that has motivated me recently is that string theory
has a large (as yet unknown) group of duality symmetries, subgroups of which
become visible for various compactifications.  If this picture is correct,
determining the group would constitute an important step towards understanding
the theory.  Since the visible groups become larger when more dimensions are
compactified, that seems to be a fruitful direction to go.  In classical
effective field theories, the duality groups are non-compact Lie groups,
realized nonlinearly by scalar fields on a symmetric space.  Quantum effects
break the symmetry to a discrete subgroup, which can often be described as the
restriction to matrices with integer entries.  A first step, however, is to
identify the continuous classical symmetry groups.  At the level of effective
field theory, it appears that generically they are finite-dimensional Lie
groups for $D \geq 3$, affine Lie algebras for $D = 2$, and hyperbolic Lie
algebras for $D = 1$.
Understanding the hyperbolic symmetries in $D = 1$ could
be very interesting.  As a more modest beginning, I have been exploring the
affine algebras for $D = 2$.  As discussed in my previous paper \cite{jhsa},
this is a
subject with a long history, so much of the relevant work has already been
done \cite{dolan,julia,breitenlohner,nicolai}. (A more complete list
of references to earlier work is given in \cite{jhsa}.)
Related discussions
of string theory dualities in two dimensions have been given recently
by a number of authors \cite{bakas,maharana,senc,kumar}.
Sen's work is the most far-reaching, as he described the discrete
duality subgroup of affine O(8,24) that occurs for the heterotic
string toroidally compactified to two dimensions.

The first paper explored the affine Lie algebra symmetries
of principal chiral models (PCM's) and symmetric space models (SSM's) in
flat two-dimensional space-time.
The purpose of this paper is to extend those results to
include coupling to gravity, which is a necessary step for eventual
application to string theory.  Many of the results described here
have been obtained previously in the cited references. However, our
derivation of the symmetry algebra is made more transparent than
previous ones (in my opinion)
by the use of a convenient contour integral representation.
Focusing on field transformations rather than Poisson brackets
also simplifies the study of the algebra, though it prevents us from
deriving the classical central extension
of the algebra found by previous authors. The only specific point
(already discussed in \cite{jhsa}) on which there seems to be a disagreement
with prior work concerns the precise description of the affine algebra
that occurs in symmetric space models.

The paper is organized as follows. Section 2 summarizes the results obtained
previously in flat space-time.  Section 3 describes the generalization of the
PCM results when gravity is included.  The main conclusion is that even though
the symmetry transformations depend on the gravitational background
(which is described by a solution of the two-dimensional wave equation),
the affine symmetry algebra remains unchanged
from flat space.  However, we are unable to generalize the Virasoro symmetry of
the flat space theory to the theory with gravity.  Section 4 describes SSM's
coupled to gravity.   There, too, we find the same symmetry algebra as in flat
space-time.

\sectiono{Summary of Previous Results}

\subsection{Principal Chiral Models}

Principal chiral models (PCM's) are based on fields $g(x)$ that map space-time
into a group manifold $G$, which we may assume to be compact.  Even though
these models are not directly relevant to the string theory and supergravity
applications that we have in mind, they serve as a good warm-up exercise, as
well as being of some interest in their own right.  Symmetric space models,
which are relevant, share many of the same features, but are a little more
complicated.  They will be described in Section 2.3.

The classical theory of PCM's, in any dimension, is defined by the lagrangian
\begin{equation}
{\cal L} = \eta^{\mu\nu} tr (A_\mu A_\nu),
\label{lagrangian}
\end{equation}
where the connection $A_\mu$ is defined in terms of the group variables by
\begin{equation}
A_\mu = g^{-1} \partial_\mu g = \sum A_\mu^i T_i.
\label{AinPCM}
\end{equation}
Here $\eta^{\mu\nu}$ denotes the Minkowski metric for flat space-time, and the
$T_i$ are the generators of the Lie algebra,
\begin{equation}
[T_i,T_j] = f_{ij}{}^k T_k.
\end{equation}
They may be taken to be matrices in any convenient representation.  The
classical equation of motion is derived by letting $\delta g$ be an arbitrary
infinitesimal variation of $g$ for which $\eta = g^{-1}\delta g$
belongs to the Lie algebra ${\cal G}$.  Under this variation
\begin{equation}
\delta A_\mu = D_\mu \eta = \partial_\mu \eta + [A_\mu, \eta],
\end{equation}
and the classical equation of motion is
\begin{equation}
\partial_\mu A^\mu = 0,
\end{equation}
as is well-known.  Since $A_{\mu}$ is pure gauge, the Bianchi identity is
\begin{equation}
F_{\mu\nu} = \partial_\mu A_\nu - \partial_\nu A_\mu + [A_\mu, A_\nu] = 0.
\end{equation}

The PCM in any dimension has manifest global $G\times G$ symmetry corresponding
to left and right group multiplication.  Remarkably, in two dimensions this is
just a small subgroup of a much larger group of ``hidden'' symmetries.  To
describe how they arise, it is convenient to introduce light-cone coordinates
\begin{equation}
x^\pm = x^0 \pm x^1,\quad \partial_\pm =
{1\over 2} (\partial_0 \pm \partial_1).
\end{equation}
Expressed in terms of these coordinates, the equation of motion and Bianchi
identity take the forms
\begin{equation}
\partial_{\mu} A^{\mu} = \partial_+ A_- + \partial_- A_+ = 0
\label{conservation}
\end{equation}
\begin{equation}
F_{+-} = \partial_+ A_- - \partial_- A_+ + [A_+, A_-] = 0.
\label{flat}
\end{equation}

A standard technique (sometimes called the ``inverse scattering method'') for
discovering the ``hidden symmetries'' of integrable models, such as a PCM in
two dimensions, begins by considering a pair of linear differential equations,
known as a Lax pair.  In the present context the appropriate equations are
\begin{equation}
(\partial_+ + \alpha_+ A_+) X = 0 \quad {\rm and} \quad
(\partial_- + \alpha_- A_-) X = 0,
\label{laxpair}
\end{equation}
where $\alpha_\pm$ are constants.
These equations are compatible, as a consequence
of eqs. (\ref{conservation}) and (\ref{flat}), provided that
\begin{equation}
\alpha_+ + \alpha_- = 2 \alpha_+ \alpha_-.
\end{equation}
It is convenient to write the solutions to this equation in terms of a
``spectral parameter'' $t$ in the form
\begin{equation}
\alpha_+ = {t\over t - 1} , \quad \alpha_- = {t\over t + 1}.
\label{spectral}
\end{equation}

The variable $X$ in eq. (\ref{laxpair})
is a group-valued function of the space-time
coordinate, as well as the spectral parameter.  The integration constant can be
fixed by requiring that $X$ reduces to the identity element of the group at a
``base point'' $x_0^\mu$.  A formal solution to
eqs. (\ref{laxpair}) is then given by a path-ordered exponential
\begin{equation}
X (x, t) = P \exp \Big\{ - \int_{x_{0}}^x (\alpha_+ A_+ dy^+
+ \alpha_- A_- dy^-)\Big\},
\end{equation}
where the path ordering has $x$ on the left and $x_0$ on the right.  The
integral is independent of the contour provided the space-time is simply
connected.  This is the case, since we are assuming a flat Minkowski
space-time.  If one were to choose a circular spatial dimension instead, the
multivaluedness of $X$ would raise new issues, which we will not consider here.
Note that $X$ is group-valued for any real $t$, except for
the singular points $t = \pm 1$.

The next step is to consider the variation
\begin{equation}
g^{-1} \delta g = \eta (\epsilon, t) = X(t) \epsilon X(t)^{-1},
\label{deltag}
\end{equation}
where $\epsilon = \sum \epsilon^i T_i$ and $\epsilon^i$ are infinitesimal
constants.  The claim is that the variation $\delta(\epsilon, t) g = g\eta$
preserves the equation of motion $\partial \cdot A = 0$ and, therefore,
describes symmetries of the classical theory.  To show this, one simply notes
that the Lax  pair implies that
\begin{equation}
\delta A_\pm = D_\pm \eta = \partial_\pm \eta + [A_\pm, \eta] = \pm {1\over t}
\partial_\pm \eta,
\end{equation}
and, therefore, $\partial \cdot (\delta A) = 0$ as required.

To compute the commutator of two infinitesimal symmetry transformations
the key identity that we require is
\begin{equation}
\delta_1 X_2 = {t_2\over t_1 - t_2} (\eta_1 X_2 - X_2 \epsilon_1),
\label{dXflat}
\end{equation}
where $\delta_i = \delta(\epsilon_i, t_i)$,
$\eta_i = \eta(\epsilon_i, t_i)$, and $X_i =X(t_i)$.
Identities such as this are used frequently in this work.  The method of proof
is to show that both sides of the equation satisfy the same
pair of linear differential equations
(obtained by varying the Lax pair) and the boundary condition
that $\delta X$ vanishes at $x_0$.
Using eqs. (\ref{dXflat}) and (\ref{deltag}), it is easy to compute
$[\delta_1,\delta_2]g$ and derive
\begin{equation}
[\delta (\epsilon_1, t_1), \delta (\epsilon_2, t_2)] = {t_1 \delta
(\epsilon_{12}, t_1) - t_2 \delta (\epsilon_{12}, t_2)\over t_1 - t_2},
\label{talgebra}
\end{equation}
where
\begin{equation}
\epsilon_{12} = [\epsilon_1, \epsilon_2] = f_{ij}{}^k
\epsilon_1^i \epsilon_2^j T_k.
\end{equation}

In order to understand the relationship between the
algebra (\ref{talgebra}) and the affine
algebra associated with the group $G$, we need to extract some sort of mode
expansion from the dependence on
the parameter $t$.  The standard approach in the
literature is to do a power series expansion in $t$, $\delta (\epsilon, t) =
\sum_{n = 0}^\infty \delta_n (\epsilon) t^n$, identifying the $\delta_n
(\epsilon)$ as distinct symmetry transformations.  This gives half of
an affine Lie algebra:
\begin{equation}
[\delta_m (\epsilon_1), \delta_n (\epsilon_2)] = \delta_{m + n} (\epsilon_{12})
\quad m,n \geq 0 .
\end{equation}
Actually, $\delta (\epsilon, t)$ contains more information than is extracted in
this way, and in Ref. \cite{jhsa}
I found a nice way to reveal it.  The idea is to
define variations $\Delta_n (\epsilon) g$ for all integers $n$ by the contour
integral
\begin{equation}
\Delta_n (\epsilon) g = \int_{\cal C} {dt\over 2\pi i} t^{-n -1}
\delta (\epsilon, t) g  \quad  n \in \ZZ,
\label{PCMcontour}
\end{equation}
where the contour ${\cal C} = {\cal C}_+ + {\cal C}_-$ and ${\cal C}_\pm$ are
small clockwise circles about $t = \pm 1$.  By distorting contours it is easy
to show that $\Delta_n (\epsilon) = \delta_n (\epsilon)$ for $n > 0$,
a result that arises entirely from the pole at $t=0$.  The
negative integers $n$ are given entirely by the pole at $t = \infty$.  In other
words, they correspond to the coefficients in a series expansion in inverse
powers of $t$.  $\Delta_0$ receives contributions from
poles at both $t = 0$ and $t =
\infty$.  (Explicitly, $\Delta_0 (\epsilon) g = [g, \epsilon]$.)  Because
$g^{-1} \Delta_n g$ can be related to such series expansions, it is clear that
it is Lie-algebra valued.\footnote{If one tried to define further symmetries
corresponding to the contours ${\cal C}_\pm$ separately or by allowing $n$ to
be non-integer, the transformations defined in this way would also appear to
preserve $\partial \cdot A = 0$.  However, these could fail to be honest
symmetries because $g^{-1} \delta g$ might not be Lie-algebra valued.}

Using the definition (\ref{PCMcontour}) and the commutator (\ref{talgebra}),
it is an easy application of
Cauchy's theorem to  deduce the affine Lie algebra (without center)
\begin{equation}
[\Delta_m (\epsilon_1), \Delta_n (\epsilon_2)] = \Delta_{m+n} (\epsilon_{12})
\qquad m,n \in {\ZZ}.
\end{equation}
Equivalently, in terms of charges, we have
\begin{equation}
[J_m^i, J_n^j] = f^{ij}{}_k J_{m+n}^k.
\end{equation}

\subsection{Virasoro symmetries}

Having found affine Lie algebra symmetries for classical PCM's, it is plausible
that they should also have Virasoro symmetries. Modulo an
interesting detail, this is indeed the case.  Since the infinitesimal parameter
in this case is not Lie-algebra valued, it can be omitted without ambiguity.
With this understanding, the Virasoro transformation is
\begin{equation}
\delta^V (t) g =  g ((t^2 - 1) \dot X(t) X(t)^{-1} + I),
\label{virsym}
\end{equation}
where the dot denotes a $t$ derivative and
\begin{equation}
I = \dot X(0) = \int_{x_{0}}^x (A_+ dy^+ - A_- dy^-).
\end{equation}
This is also an invariance of the equation of motion $\partial \cdot A =
0$.  We can extract modes $\delta_n^V$, for all integers $n$, by the same
contour integral definition used above
\begin{equation}
\delta_n^V g = \int_{\cal C} {dt\over 2\pi i} t^{-n -1} \delta^V (t) g.
\end{equation}
Again, contour deformations give pole contributions at $t = 0$ and $t = \infty$
only, and therefore, one sees that $g^{-1} \delta_n^V g$ is Lie-algebra valued.

The analysis of the algebra proceeds in the same way as for the affine
symmetry algebra,
though the formulas are quite a bit more complicated.  For example, commuting
a Virasoro symmetry transformation with an affine algebra symmetry
transformation gives
\[[\delta^V (t_1), \delta (\epsilon, t_2)] g =  \Big({1\over t_2}
(\delta (\epsilon, 0) - \delta (\epsilon, t_2))
+ {t_2 (t_1^2 - 1)\over (t_1 - t_2)^2} (\delta (\epsilon, t_1)
- \delta (\epsilon, t_2))\]
\begin{equation}
+ {t_1 (1 - t_2^2)\over t_1 - t_2} ~ {\partial\over\partial t_2} \delta
(\epsilon, t_2) \Big) g .
\label{PCMvir}
\end{equation}
Using this equation and the contour integral definitions,
one finds after an
integration by parts and use of Cauchy's theorem that
\begin{equation}
[\delta_{m}^V, \Delta_{n} (\epsilon)] g = n \int_{\cal C} {dt\over
2\pi i} t^{-m -n -2} (t^2 -1) \delta (\epsilon, t) g.
\label{vircomm}
\end{equation}

Now let us re-express the algebra in terms of charges $J_n^i$ (as before) and
$K_m$ (corresponding to $\delta_m^V$).
In this notation, eq. (\ref{vircomm}) becomes
\begin{equation}
[K_{m}, J^i_{n}] = n (J^i_{m + n-1} - J^i_{m + n + 1}).
\label{KJalg}
\end{equation}
This formula is to be contrasted with what one would expect for the usual
Virasoro generators $L_n$
\begin{equation}
[L_{m}, J^i_{n}] = - n J^i_{m + n}.
\end{equation}
Comparing equations, we see that we can make contact with
the usual (centerless)
Virasoro algebra if we identify
\begin{equation}
K_n = L_{n + 1} - L_{n - 1}.
\label{KfromL}
\end{equation}
However, it should be stressed that we have only defined the differences $K_n$
and not the individual $L_n$'s.  Still, this identification is useful since it
tells us that
\begin{equation}
[K_m, K_n] = (m - n) (K_{m+n+1} - K_{m+n-1}).
\end{equation}

Let us see what happens if we try to construct the $L_n$'s.  The easiest
approach is to define $K(\sigma) = \sum_{-\infty}^\infty K_n e^{in\sigma}$ and
$T(\sigma) = \sum_{-\infty}^\infty L_n e^{in\sigma}.$
Then eq. (\ref{KfromL}) implies that
\begin{equation}
T(\sigma) = {i\over 2} ~ {K(\sigma)\over\sin \sigma}.
\label{stress}
\end{equation}
The remarkable fact is that $K(\sigma)$ does not vanish at $\sigma = 0$ and
$\sigma = \pi$.  Therefore, $T (\sigma)$ diverges at these points and the
individual $L_n$'s do not exist.  The integrals that would define them are
logarithmically divergent.

\subsection{Symmetric Space Models}

An interesting class of integrable two-dimensional models consists of theories
whose fields map the space-time into a symmetric space.  Let $G$ be a simple
group and $H$ a subgroup of $G$.  Then the Lie algebra ${\cal G}$ can be
decomposed into the Lie algebra ${\cal H}$ and its orthogonal complement ${\cal
K}$, which contains the generators of the coset $G/H$.  The coset space $G/H$
is called a symmetric space if $[{\cal K, K}]\subset {\cal H}$, in other words
the commutators of elements of ${\cal K}$ belong to ${\cal H}$.  The examples
that arise in string theory and supergravity are non-compact symmetric space
models (SSM's).  For such models,
$G$ is a non-compact Lie group and $H$ is its maximal compact subgroup.  The
generators of ${\cal H}$ are antihermitian and those of ${\cal K}$ are
hermitian.  Therefore, since the commutator of two hermitian matrices is
antihermitian, $[{\cal K, K}]\subset {\cal H}$ and $G/H$ is a (non-compact)
symmetric space.

Symmetric space models can be formulated starting with arbitrary $G$-valued
fields, $g(x)$, like those of PCM's.  To construct an SSM, we associate local
$H$ symmetry with left multiplication and global $G$ symmetry with right
multiplication.  Thus, we require invariance under infinitesimal
transformations of the form
\begin{equation}
\delta g = - h (x) g +
g \epsilon \quad h\in {\cal H}, \  \epsilon\in {\cal G}.
\end{equation}
The local symmetry effectively removes $H$ degrees of freedom so that only
those of the coset remain.  The next step is to define
\begin{equation}
A_\mu = M^{-1} \partial_\mu M,
\label{Aform}
\end{equation}
where
\begin{equation}
M = g^\dagger g.
\label{Mform}
\end{equation}
Note that $g$ and $M$ are analogous to a vielbein and metric in general
relativity.  $M$, which is invariant under local $H$ transformations,
parametrizes the symmetric space $G/H$ without extra degrees
of freedom.  In the case of a compact SSM the factor
$g^\dagger$ in the definition of
$M$ must be generalized to a quantity $\tilde{g}$, which is described in Ref.
\cite{jhsa}.  Since $A_\mu$ is pure gauge, its field strength vanishes
\begin{equation}
F_{\mu\nu} = \partial_\mu A_\nu - \partial_\nu A_\mu + [A_\mu, A_\nu] = 0.
\end{equation}
The lagrangian is ${\cal L} = {\rm tr} (A^\mu A_\mu)$
and the classical equation of motion is
\begin{equation}
\partial^{\mu} A_{\mu} = 0.
\end{equation}
These formulas look the same as for PCM's, but $A_\mu$ is given in terms of
$g(x)$ by a completely different formula
(eqs. (\ref{Aform}) and (\ref{Mform}) instead of eq. (\ref{AinPCM}) ).

In two dimensions we once again have the Bianchi identity  $F_{+-} = 0$ and the
equation of motion $\partial_+ A_- + \partial_- A_+ = 0$.  Therefore, it is
natural to investigate whether the formulas that gave rise to symmetries of
PCM's also gives rise to symmetries in this case.  With this motivation, we
once again form the Lax pair of equations
\begin{equation}
(\partial_\pm + \alpha_\pm A_\pm) X = 0,
\end{equation}
and note that they are compatible if we write $\alpha_\pm$ in terms of a
spectral parameter $t$ as in eq. (\ref{spectral}).
Then the solution is given by the contour-independent integral
\begin{equation}
X(t) = P \exp \Big(- \int_{x_{0}}^x (\alpha_+ A_+ dy^+
+\alpha_- A_- dy^-) \Big),
\end{equation}
as before.  The obvious guess is that, just as for PCM's, the hidden
symmetry is described by
\begin{equation}
\delta g = g X(t) \epsilon X(t)^{-1} .
\end{equation}
This turns out to be correct.  Under an arbitrary infinitesimal variation
$g^{-1} \delta g = \eta (x) \in {\cal G}$, we have
\begin{equation}
\delta M = \eta^\dagger M + M \eta ,
\end{equation}
which implies that
\begin{equation}
\delta A_\mu = D_\mu \eta + D_\mu (M^{-1} \eta^\dagger M).
\label{dASSM}
\end{equation}
The first term is the same as for a PCM, but the second one is new.  The
symmetry requires that $\partial^\mu (\delta A_\mu) = 0$, when we substitute
$\eta = X \epsilon X^{-1}$.  The vanishing of the contribution from the first
term in eq. (\ref{dASSM}) is identical to the PCM case.
The second term in eq. (\ref{dASSM})
also has a vanishing divergence (for $\eta = X \epsilon X^{-1}$).

Next, we wish to study the algebra of these symmetry transformations.  The
first step is to derive a suitable generalization of eq. (\ref{dXflat}),
which is
\begin{equation}
\delta_1 X_2 = {t_2\over t_1 - t_2} (\eta_1 X_2 - X_2 \epsilon_1) +
{t_1 t_2\over 1- t_1 t_2} (M^{-1} \eta_1^\dagger MX_{2}
-X_2 M_0^{-1}\epsilon_1^{\dagger} M_0), \label{eq:dxssm}
\end{equation}
where $M_0 = M(x_0)$.
The first term is the one we found for PCM's.  The second term, which is new,
is required to compensate for the extra piece of $\delta A_\mu$ in
eq. (\ref{dASSM}) that occurs for SSM's.
The commutator is then found to be
\begin{equation}
[\delta(\epsilon_1, t_1), \delta(\epsilon_2, t_2)]g =
{t_1 \delta (\epsilon_{12}, t_1) -
t_2 \delta (\epsilon_{12}, t_2)\over t_1 - t_2}g +\delta^{\prime}g
+ \delta^{\prime\prime}g ,
\label{SSMcomm}
\end{equation}
where the first term is the same as we found for PCM's, but there are two
additional pieces.  The $\delta' g$ term is a local ${\cal H}$ transformation
of the form $h_{12}(x) g$, with
\begin{equation}
h_{12} (x) = {t_1 t_2\over 1 - t_1 t_2 }
\Big[(g^\dagger)^{-1} \eta_1^\dagger M \eta_2 g^{-1}
+ g \eta_1 M^{-1} \eta_2^\dagger g^\dagger\Big] - {\rm h.c.},
\label{htrans}
\end{equation}
which is a symmetry of the theory.  It is trivial in its
action on $M = g^\dagger g$, which is all that appears in ${\cal L}$. The
$\delta^{\prime\prime} g$ term is given by
\begin{equation}
\delta^{\prime\prime}g = {t_1 t_2\over 1 - t_1 t_2 }
\left( \delta(\epsilon'_{21},t_1) - \delta(\epsilon'_{12}, t_2) \right)g,
\end{equation}
where
\begin{equation}
\epsilon'_{12} = M_0^{-1} \epsilon_1^{\dagger} M_0 \epsilon_2
-\epsilon_2 M_0^{-1}\epsilon_1^{\dagger} M_0.
\end{equation}

As in the PCM, we define modes by contour integrals of the form given in eq.
(\ref{PCMcontour}), and associate charges $J_n^i$
to the transformation $\Delta_n
(\epsilon)$.  These can be converted to ``currents'' $J^i (\sigma) = \sum
e^{in\sigma} J_n^i$.  In the case of an SSM, there are two distinct classes of
currents, those belonging to ${\cal H}$ and those belonging to ${\cal K}$.  As
Ref. \cite{jhsa} shows in detail, the significance of the
$\delta^{\prime\prime} g$
term in eq. (\ref{SSMcomm}) is that
the ${\cal H}$ currents satisfy Neumann boundary
conditions at the ends of the interval $0 \leq \sigma \leq \pi$, while the
${\cal K}$ currents satisfy Dirichlet boundary conditions at the two ends
\begin{equation}
J^{i\prime} (0) = J^{i\prime} (\pi) =0 \quad{\rm for}
\quad J^i \in {\cal H}
\end{equation}
\begin{equation}
J^i (0) =J^i (\pi) = 0 \quad{\rm for}\quad J^i \in {\cal K}.
\end{equation}
As a result, $J_n^i = J_{-n}^i$ for ${\cal H}$ charges and
$J_n^i = - J_{-n}^i$ for
${\cal K}$ charges.
In terms of the modes, the affine symmetry algebra on the line segment
$0 \leq \sigma \leq \pi$ then implies that
\begin{equation}
[J_m^i, J_n^j] = f^{ij}{}_k (J_{m + n}^k + J_{m-n}^k) \quad{\rm for}\quad
J_n^j \in {\cal H}
\end{equation}
\begin{equation}
[J_m^i, J_n^j] = f^{ij}{}_k (J_{m+n}^k - J_{m-n}^k) \quad{\rm for}
\quad J_n^j \in {\cal K}.
\end{equation}
I propose to call this kind of an affine Lie algebra $\hat G_H$.

The Virasoro symmetries of PCM's also generalize to SSM's.  The natural
guess is that, just as for the affine algebra symmetry, the same formula will
describe the symmetry in this case, namely
\begin{equation}
\delta^V (t) g =
g \Big((t^2 - 1) \dot X (t) X (t)^{-1} + I\Big).
\end{equation}
This turns out to be correct, but once again the algebra differs from that of
PCM's.  We find that
\begin{equation}
[\delta^V (t_1), \delta(\epsilon, t_2)] g =
\delta g + \delta' g + \delta^{\prime\prime} g,
\end{equation}
where $\delta g$ is the PCM result given in
eq. (\ref{PCMvir}).  The $\delta' g$ is a
local ${\cal H}$ transformation and $\delta^{\prime\prime} g$ contains new
terms.  (The formulas are given in
Ref. \cite{jhsa}.)  The crucial question becomes
what $\delta^{\prime\prime} g$ contributes to $[\delta_m^V, \delta_n
(\epsilon)]g$, when we insert it into the appropriate contour integrals, or,
equivalently, what it contributes to $[K_m, J_n^i]$.  The result is
\begin{equation}
[K_m, J_n^i] = n (J_{m+n-1}^i - J_{m+n+1}^i - J_{n-m+1}^i + J_{n-m-1}^i).
\label{newKJalg}
\end{equation}
The first two terms are the PCM result of
eq. (\ref{KJalg}), while the last two terms
are the new contribution arising from $\delta^{\prime\prime} g$.

After our experience with the affine Lie algebra symmetry,
the interpretation of
the result (\ref{newKJalg}) is evident.
The generators $K_m$ satisfy the restrictions $K_m
= K_{-m}$, just like the ${\cal H}$ currents.  In other words, $K(\sigma)$
satisfies Neumann boundary conditions at the ends of the interval $0 \leq
\sigma \leq \pi$.  Just as for PCM's, one can define a stress tensor
by eq. (\ref{stress}),
which satisfies the standard stress tensor algebra.
As before, it is singular at
$\sigma=0$ and $\sigma=\pi$, so that modes $L_m$ do not exist.

\sectiono{Principal Chiral Models Coupled to Gravity}

\subsection{Formulation of the theory}

Coupling a principal chiral model to gravity is completely straightforward
at the classical level for $D>2$.  One simply generalizes the
lagrangian in (\ref{lagrangian}) to
\begin{equation}
{\cal L} = \sqrt{-h} (R(h) + h^{\mu\nu} tr (A_\mu A_\nu)).
\label{eq:lagrangian}
\end{equation}
We use $h_{\mu\nu}$ for the space-time metric, since the symbol $g$ has
already been used for the group variable.  As before, $A_\mu = g^{-1}
\partial_\mu g$, and $R(h)$ is the scalar curvature, of course.  Again,
normalization factors are omitted, since our considerations are classical.  Our
main interest is in two dimensions, and it is important that we choose the
right theory in that case
--- the one that is relevant to string theory.
The above formula in $D=2$ is not what we want.
Instead, the theory that we shall consider is obtained by starting with
eq. (\ref{eq:lagrangian})
in three dimensions and doing a dimensional reduction to two dimensions.  This
means simply dropping the dependence of the fields on one of the spatial
coordinates, which is a consistent truncation.

To do the dimensional reduction from three dimensions to two dimensions, it is
convenient to decompose the dreibein $\hat e_\mu^a$ in terms of a zweibein
$e_\mu^a$, a vector $B_\mu$, and a scalar $\rho$ as follows
\begin{equation}
\hat e_\mu^a = \left(\begin{array}{cc}
e_\mu^a & \sqrt{\rho} B_\mu \\
0 & \sqrt{\rho} \end{array}\right).
\end{equation}
The zeros are obtained by gauge fixing the local Lorentz transformations
between the dimension we are dropping and the two we are keeping.  In terms of
the metric, this formula corresponds to
\begin{equation}
\hat h = \hat e \eta \hat e^T = \left(\begin{array}{cc}
h_{\mu\nu} + \rho B_\mu B_\nu & \rho B_\mu \\
\rho B_\nu & \rho \end{array} \right).
\end{equation}
Making these substitutions in the lagrangian, dropping all derivatives in the
$x^2$ direction, and defining $G_{\mu\nu} = \partial_\mu B_\nu - \partial_\nu
B_\mu$, leads to
\begin{equation}
{\cal L}_2^{(0)} = \sqrt{-h} \rho (R(h) - {1\over 2} \rho^2 G_{\mu\nu}
G^{\mu\nu} + h^{\mu\nu} tr (A_\mu A_\nu)),
\end{equation}
a result that has been obtained by many authors.  By considering the $B_\mu$
field equation it is easy to convince oneself that this field has no effect
on the theory.  The $G^2$ term can simply be dropped, which is what
I will do. (In two dimensions it is possible to set $G_{\mu\nu}$
proportional to $\epsilon_{\mu\nu}$, which gives rise to a cosmological
constant. We will not do that, however.)

Because of the work on string theory, a great deal is
known about two-dimensional gravity theories.  One lesson
is that the story is particularly simple when there is conformal symmetry.
Indeed that is an important requirement in string theory.  Since the theory
being considered here is regarded as a target-space theory, rather than a
world-sheet theory,
there is no reason that we should require conformal symmetry.
Indeed we do not have it.  Specifically, if we rescale the metric by
$h_{\mu\nu}
\rightarrow e^\phi h_{\mu\nu}$, one finds that (up to a total derivative)
\begin{equation}
{\cal L}_2^{(0)} \rightarrow  {\cal L}_2 = {\cal L}_2^{(0)} + \sqrt{-h}
h^{\mu\nu} \partial_\mu \rho \partial_\nu \phi.
\end{equation}
We can take $\phi$ to be an additional independent field and
define the theory to be given by  ${\cal L}_2$.  Then we do have
local Weyl invariance under
\begin{equation}
h_{\mu\nu} \rightarrow e^\lambda h_{\mu\nu}, \quad \phi \rightarrow \phi -
\lambda.
\end{equation}
This has the practical consequence that the energy--momentum tensor
$T_{\mu\nu}$
is traceless.

Let us now form all the equations of motion.  After forming them we can use the
diffeomorphism and Weyl symmetries to choose the gauge $h_{\mu\nu} =
\eta_{\mu\nu}$.  The equations of motion in this gauge become
\begin{equation}
\partial_+ (\rho A_-) + \partial_- (\rho A_+) = 0
\label{first}
\end{equation}
\begin{equation}
\partial_+ \partial_- \rho = 0
\end{equation}
\begin{equation}
\partial_+ \partial_- \phi = {\rm tr} (A_+ A_-)
\end{equation}
\begin{equation}
T_{++} = \rho {\rm tr}
(A_+ A_+) + \partial_+ \phi \partial_+ \rho + c \partial_+^2
\rho = 0
\end{equation}
\begin{equation}
T_{--} = \rho {\rm tr} (A_- A_-) + \partial_- \phi \partial_- \rho + c
\partial_-^2
\rho = 0 ,
\label{last}
\end{equation}
where $c$ is a constant. Equation (\ref{first}) is sometimes called
the Ernst equation. We begin by solving the 2D wave equation for $\rho$:
\begin{equation}
\rho (x) = \rho^+ (x^+) + \rho^- (x^-).
\end{equation}
{}From now on $\rho^+ (x^+)$ and $\rho^- (x^-)$ will be regarded as arbitrary
given functions that describe a ``fixed gravitational background.''  One could
redefine coordinates by setting $\rho^+ (x^+) = x^+$ and $\rho^- (x^-) = x^-$,
for example, without loss of generality.  However, there is no need to do this,
and I prefer to keep $\rho ^+$ and $\rho^-$ arbitrary.
The constant $c$, which appears in $T_{++}$ and $T_{--}$, is of no consequence,
because it can be absorbed in a redefinition
\begin{equation}
\phi \rightarrow \phi + c \log (\partial_+ \rho \partial_- \rho).
\end{equation}
Therefore, we will set it to zero.

The last three of the equations of motion,
the ones that involve $\phi$, can be solved explicitly by giving a formula for
$\phi$ provided one uses the  other two equations.  To see this, let us first
solve $T_{++} = 0$ for $\partial_+ \phi$:
\begin{equation}
\partial_+ \phi = - {\rho\over \partial_+ \rho} {\rm tr} (A_+ A_+).
\end{equation}
Before attempting to integrate this, let us differentiate it with respect
$x^-$, using the identity
\begin{equation}
\partial_- A_+ = {1\over 2} [A_+, A_-] - {1\over 2\rho} (\partial_+ \rho A_- +
\partial_- \rho A_+),
\end{equation}
which is obtained by combining $F_{+-} = 0$ and $\partial_\mu (\rho A^\mu) =
0$.  This immediately leads to
\begin{equation}
\partial_+ \partial_- \phi = {\rm tr} (A_+ A_-),
\end{equation}
which is therefore not an independent equation.  The same result is obtained
starting from $T_{--} = 0$.  Therefore, the $T_{++}$ and $T_{--}$ equations are
compatible and can be integrated to give
\begin{equation}
\phi(x) = \phi(x_0) - \int_{x_{0}}^x \rho \left[{{\rm tr}
(A_+ A_+)\over \partial_+ \rho} dy^+ + {{\rm tr}
(A_- A_-)\over \partial_- \rho} dy^-\right] ,
\label{eq:phiformula}
\end{equation}
which is contour independent, when the equation of motion is used.  Thus,
aside from its value at one point,
$\phi$ is given in terms of $g$ and $\rho$.

\subsection{The symmetry transformations}

Our goal is to generalize the hidden symmetries
that we found for PCM's in 2D flat
space to PCM's coupled to 2D gravity.  To do this, we follow the same steps as
before.  First we seek a Lax pair
\begin{equation}
(\partial_{\pm} + \alpha_{\pm} A_{\pm}) X = 0,
\end{equation}
whose consistency follows from the equation of motion $\partial_\mu (\rho
A^\mu) = 0$ and the Bianchi identity $F_{+-} = 0$.  As before, a necessary
condition is that ${1\over\alpha^+} + {1\over \alpha^-} = 2$, so once again we
write
\begin{equation}
\alpha_+ = {\tau\over\tau - 1} , \qquad \alpha_- = {\tau\over \tau + 1} .
\label{alphas}
\end{equation}
We now use the symbol $\tau$ in place of $t$, which we used earlier, because
$\tau$ will turn out to be $x^\mu$ dependent, and we want to reserve the symbol
$t$ for a constant that will be defined later.

Requiring
$[\partial_+ +\alpha_+ A_+, \partial_- +
\alpha_- A_-] = 0$,
and using $\partial_\mu (\rho A^\mu) = 0$ and $F_{+-} = 0$, gives the
differential equations :
\begin{equation}
2\rho\partial_+\alpha_- = (\alpha_- - \alpha_+) \partial_+ \rho
\quad {\rm and} \quad
2\rho\partial_-\alpha_+ = (\alpha_+ - \alpha_-) \partial_- \rho
\end{equation}
Suppose we now let
\begin{equation}
\tau = {1-R\over 1+R},
\end{equation}
so that
\begin{equation}
\alpha_+ = {1\over 2} \left(1 - {1\over R}\right), \quad
\alpha_- = {1\over 2} (1 - R) .
\end{equation}
Substituting these expressions then gives the equations
\begin{equation}
R^{-1}\partial_+ R = \left(1 - R^{-2}\right) {\partial_+ \rho\over 2\rho}
\quad {\rm and} \quad
R^{-1}\partial_- R = \left(R^2 - 1\right) {\partial_- \rho\over 2\rho} .
\end{equation}
The general solution of this pair of equations is
\begin{equation}
R = \left({1 - \kappa \rho^+\over 1 + \kappa \rho^-}\right)^{1/2},
\end{equation}
where $\kappa$ is an integration constant. To define the branch of the
square root unambiguously,
we restrict attention to the neighborhood of $\kappa = 0$, avoiding branch
points, with the understanding that $R \rightarrow + 1$ as $\kappa \rightarrow
0$.
The function
\begin{equation}
\tau (\kappa, x) = {\sqrt{1 + \kappa\rho^-} - \sqrt{1 - \kappa
\rho^+}\over \sqrt{1 + \kappa\rho^-}
+ \sqrt{1 - \kappa \rho^+}},
\label{tauform}
\end{equation}
also has a series expansion in $\kappa$, whose first term is
$\tau (\kappa, x) \sim {1\over 4} \kappa \rho$.
Notice that changing the choice of sheet corresponds to
the transformation $\tau \rightarrow 1/\tau$.

Now we can generalize the symmetry transformations of $g(x)$
that were obtained in the flat-space case:
\begin{equation}
\delta (\epsilon, \kappa) g (x) = F (\kappa, x) g (x) \eta
(\epsilon, \kappa, x),
\end{equation}
where $\eta$ is defined exactly as before
\begin{equation}
\eta (\epsilon, \kappa, x) = X (\kappa, x) \epsilon X (\kappa, x)^{-1}
\label{etaeqn}
\end{equation}
\begin{equation}
X (\kappa,x) = P\exp \left\{ - \int_{x_{0}}^x
(\alpha_+ A_+ dy^+ + \alpha_- A_- dy^-)\right\}.
\label{Xeqn}
\end{equation}
There are two significant differences from the flat-space case.  First,
$\alpha^\pm$, as defined by eqs. (\ref{alphas}) and (\ref{tauform}),
are functions of $y^\mu$ and $\kappa$.
Second, we allow for an extra
factor $F(\kappa,x)$ in the variation.  This factor is required to be a scalar
function, unlike $g$ and $\eta$ which are matrices. Thus, so long as $F$ is
real, $g^{-1} \delta g$ is Lie-algebra valued.

The question now is whether there is a choice of $F(\kappa,x)$ such that the
equation of motion $\partial_\mu (\rho A^\mu) = 0$
is preserved under the variation
$\delta g$ given above.  To examine this, we first note that
\begin{equation}
\delta (\epsilon, \kappa) A_\pm = D_\pm (F\eta) = \pm{F\over \tau}
\partial_{\pm} \eta + \eta \partial_{\pm} F.
\end{equation}
Requiring that $\partial_\mu (\rho \delta A^\mu) =0$ then gives the conditions
\begin{equation}
\rho \partial_\pm F = \pm \partial_\pm ( F \rho /\tau).
\label{taudown}
\end{equation}
These equations
imply that, up to a multiplicative factor, $F$ is given by
\begin{equation}
f (\kappa, x) = [(1 - \kappa \rho^+) (1 + \kappa\rho^-)]^{-1/2}
\label{fform}
\end{equation}
which is again unambiguous in the neighborhood of $\kappa = 0$.  The choice of
normalization that will turn out to be most convenient is
\begin{equation}
F(\kappa, x) = {f(\kappa, x)\over f(\kappa, x_0)}.
\label{Fform}
\end{equation}
Specifically, it follows that
\begin{equation}
\delta A_\pm = D_\pm (F\eta) = \pm {1\over\rho}
\partial_\pm ( F \rho\eta / \tau),
\end{equation}
so that the conserved charges are given by
\begin{equation}
Q(\epsilon,t) = \int_{-\infty}^\infty \rho D_0 (F\eta) dx^1 =
\left({\rho F\over \tau} X \epsilon X^{-1}\right)\Bigg|_{- \infty}^\infty.
\end{equation}

\subsection{The symmetry algebra}

To derive the commutator of two transformations $[\delta_1, \delta_2] =
[\delta(\epsilon_1, \kappa_1), \delta
(\epsilon_2, \kappa_2)]$, we need to generalize the
formula for $\delta_1 X_2$ in the flat-space theory.  Guided by
what we found in eq. (\ref{dXflat}) in that case, let us try
\begin{equation}
\delta_1 X_2 = \zeta_{12} \eta_1 X_2 + \lambda_{12} X_2 \epsilon_1,
\label{dXPCM}
\end{equation}
where $\zeta_{12}$ and $\lambda_{12}$ are unknown functions to be
determined.  The important point is that, like $F$, they are not matrices.  The
required conditions are obtained by varying the equations $(\partial_\pm +
\alpha_\pm A_\pm) X = 0$.  The resulting equations for $\zeta_{12}$ and
$\lambda_{12}$ tell us that $\lambda_{12}$ is a constant and
\begin{equation}
\zeta_{12} = {\tau_2 F_1\over \tau_1 - \tau_2}.
\end{equation}
They also require that
\begin{equation}
\partial_{\pm} \zeta_{12} + {\tau_2\over \tau_2 \mp 1} \partial_{\pm} F_1 =0,
\end{equation}
which is true for this choice of $\zeta_{12}$.
The constant $\lambda_{12}$ is determined by noting that since
$X_2 \rightarrow 1$ as $x \rightarrow x_0$, it is necessary that
$\delta_1 X_2 (x_0) = 0$.  This then implies that
\begin{equation}
\lambda_{12} = - \zeta_{12} (x_0) = {t_2\over t_2 - t_1},
\end{equation}
where we have defined
\begin{equation}
t_i = \tau (\kappa_i, x_0).
\label{tdef}
\end{equation}
Henceforth we use this formula to replace $\kappa_i$ by $t_i$.
Note that for small $\kappa_i$, $t_i \sim {1\over 4} \kappa_i \rho (x_0)$.

The algebra is now easy to derive.
The formula for $\delta_1 X_2$ implies that
\begin{equation}
\delta_1 \eta_2 = [\delta_1 X_2 \cdot X_2^{-1}, \eta_2] =
\zeta_{12} [\eta_1, \eta_2] + \lambda_{12} X_2
\epsilon_{12} X_2^{-1},
\end{equation}
where $\epsilon_{12} = [\epsilon_1, \epsilon_2]$, as before.
Now the commutator of two symmetry transformations can be evaluated using
\begin{equation}
g^{-1} [\delta_1, \delta_2] g = F_1 F_2 [\eta_1, \eta_2]
+F_2 \delta_1 \eta_2 - F_1 \delta_2 \eta_1
\end{equation}
and the identity
\begin{equation}
F_1 F_2 + F_2 \zeta_{12} + F_1 \zeta_{21} =0.
\end{equation}
Then one finds precisely the same formula as in the flat-space case, namely
\begin{equation}
[\delta (\epsilon_1, t_1), \delta (\epsilon_2, t_2)]g = {t_1 \delta
(\epsilon_{12}, t_1) - t_2 \delta (\epsilon_{12}, t_2)\over t_1 - t_2} g.
\label{gravcomm}
\end{equation}

We may expand
\begin{equation}
\delta (\epsilon, t) = \sum_{n = 0}^\infty \delta_n (\epsilon) t^n
\end{equation}
and obtain half of an affine algebra
\begin{equation}
[\delta_m (\epsilon_1), \delta_n (\epsilon_2)] g = \delta_{m+n} (\epsilon_{12})
g.
\end{equation}
However, by using a contour integral representation for the modes,
instead, a complete affine algebra can be obtained. We shall do that
shortly, but let us first consider the field $\phi(x)$.
Equation~(\ref{eq:phiformula})
implies that $\phi(x) -\phi(x_0)$ satisfies the same algebra
as $g(x)$:
\begin{equation}
[\delta (\epsilon_1, t_1), \delta (\epsilon_2, t_2)] (\phi(x) -\phi(x_0))
= {t_1 \delta (\epsilon_{12}, t_1) - t_2 \delta (\epsilon_{12}, t_2)\over t_1 -
t_2} (\phi(x) -\phi(x_0)).
\end{equation}
This still leaves the possibility that the commutator also gives rise to
a constant translation of $\phi(x)$, which would correspond to a
global conformal
rescaling of the two-dimensional metric. Such a central term
was found to occur in refs. \cite{julia,breitenlohner,nicolai}.
The fact that it is undetermined here may indicate a limitation of our
approach, which is based on classical field transformations rather
than Poisson brackets.

The contour integral construction of $\Delta_n$
presented in Section 2.1 can be generalized to the case of a 2D PCM
coupled to gravity.
The key step is to understand the analytic structure of $X(t)$.
Whereas it had isolated singularities at $t = \pm 1$ in the flat space theory,
it has branch cuts in the theory with gravity.  To see this, it is useful to
express the function $\tau (x)$ in terms of the parameter $t$ by eliminating
the parameter $\kappa$.  A little algebra gives
\begin{equation}
\kappa = {4\tau\over \rho (1 + \tau^2) + 2 \tilde{\rho} \tau} =
{4t\over \rho_0 (1 + t^2) + 2 \tilde{\rho}_0 t},
\end{equation}
where we have defined $\rho_0 = \rho(x_0)$ and
\begin{equation}
\tilde\rho(x) = \rho^+(x^+) - \rho^-(x^-).
\end{equation}
{}From this it follows that
\begin{equation}
R = {1 - \tau\over 1 + \tau} =
\left({\rho_0 (1 + t^2) + 2 \tilde{\rho}_0 t - 4 t
\rho^+\over \rho_0 (1 + t^2) + 2 \tilde{\rho}_0 t + 4 t \rho^-}\right)^{1/2}.
\end{equation}
Coordinate dependence enters this formula through the functions $\rho^\pm
(x^\pm )$.

The quantities $\alpha_+ = {1\over 2} ( 1 - R^{-1})$ and $\alpha_- = {1\over 2}
(1 - R)$, which appear inside the integral defining $X(t)$, are singular
whenever the numerator or the denominator of $R$ vanishes.  This defines branch
cuts connecting branch points at the locations where $\alpha_\pm (x_0)$ and
$\alpha_\pm (x)$ are singular.  The roots of the expression in the numerator
of $R$ are given by
\begin{equation}
n_\pm (x^+) = {2\rho^+ - \tilde{\rho}_0 \pm 2 [ (\rho^+ + \rho_0^-) (\rho^+ -
\rho_0^+)]^{1/2}\over\rho_0}
\end{equation}
and those of the denominator are
\begin{equation}
d_\pm (x^-) = {-2\rho^- - \tilde{\rho}_0 \pm 2 [(\rho^- + \rho_0^+)(\rho^- -
\rho_0^-)]^{1/2}\over\rho_0}.
\end{equation}
At the base point $x_0$ one finds that $n_\pm (x_0) = 1$ and $d_\pm (x_0) = -
1$.  Therefore, there are branch cuts connecting $t = 1$ to $t = n_\pm (x)$ and
branch cuts connecting $t = - 1$ to $t = d_\pm (x)$.

Now we can again define
\begin{equation}
\Delta_n (\epsilon) g = \int_{\cal C} {dt\over 2\pi i}
t^{-n -1} \delta (\epsilon,t) g,
\end{equation}
with ${\cal C} = {\cal C}_+ + {\cal C}_-$.
The new feature is that now ${\cal C}_+$ must enclose the branch
cuts that connect $t = 1$ to $t = n_\pm (x)$ and ${\cal C}_-$
must enclose the branch
cuts that connect $t = - 1$ to $t = d_\pm (x)$.  Once this is done,
$\Delta_n$ becomes a well-defined finite expression.  The only thing that
needs to be checked is that the contours never get pinched against the points
$t = 0$ or $t = \infty$.  In other words we must examine when $n_\pm (x)$
and $d_\pm (x)$ can vanish or diverge.  A little algebra shows that the only
way this can happen is if $\rho^+ (x^+)$ or $\rho^- (x^-)$ becomes infinite.
However, this is not allowed, at least in the finite plane, and therefore
should not be a problem.
It can now be demonstrated
that $[\Delta_{m} (\epsilon_1), \Delta_{n}
(\epsilon_2)] = \Delta_{m + n} (\epsilon_{12})$ follows from
eq. (\ref{gravcomm}) exactly as in the flat space theory.

\subsection{Virasoro symmetries}

Having found a generalization of the flat space affine algebra symmetries,
it is plausible that it should be possible to do the same for the Virasoro
symmetries.  In fact, despite considerable effort, I have been unable to find
the desired formulas.  It is unclear to me whether this failure reflects a
fundamental obstruction or a lack of ingenuity.  The discussion that follows
shows what has been achieved and the difficulties that were encountered.

Equation (\ref{virsym}), the flat space Virasoro symmetry formula, contains two
terms.
Let us begin by presenting a curved space generalization of the first term:
\begin{equation}
g^{-1} \delta^{(a)} g = (t^2 - 1) F \dot X (t) X (t)^{-1},
\end{equation}
where $t$ is the constant spectral parameter defined in
eq. (\ref{tdef}), $F$ is the
function defined in eqs. (\ref{fform}) and (\ref{Fform}),
and the dot represents a derivative with respect
to $t$.  This term gives
\begin{eqnarray}
\delta^{(a)} A_\pm &=& (t^2 - 1) D_\pm (F \dot X X^{-1})\nonumber\\
&=& (t^2 - 1) \left\{(\partial_\pm F) \dot X X^{-1} \pm {1\over\tau} F
\partial_\pm (\dot X X^{-1}) - {\dot\alpha_\pm\over\alpha_\pm}
FA_\pm\right\}\nonumber\\
&=& (t^2 - 1) \left\{ \pm {1\over\rho} \partial_\pm \left({\rho F\over\tau}
\dot X X ^{-1}\right) - {\dot\alpha_\pm\over\alpha_\pm} FA_\pm \right\},
\end{eqnarray}
where the last step uses eq. (\ref{taudown}).
The first term in the last expression does
not contribute to $\partial^\mu (\rho \delta A_\mu)$, so we can ignore it,
and focus on the second term.

The expression
\begin{equation}
\delta' A_\pm = (1 - t^2) {\dot\alpha_\pm \over \alpha_\pm} FA_\pm,
\end{equation}
can be re-expressed using the identities
\begin{equation}
{\dot\alpha_\pm\over\alpha_\pm} = {\dot\tau\over\tau (1 \mp \tau)},
\end{equation}
and
\begin{equation}
\dot\tau = {\tau\over t} F,
\end{equation}
in the form
\begin{equation}
\delta' A_\pm = {1-t^2\over t} {1\over 1 \mp \tau} F^2 A_\pm.
\end{equation}
In flat space $\tau = t$ and $F = 1$, so this reduces to $(t^{-1} \pm 1)
A_\pm$, which has divergence $\partial_- A_+ - \partial_+ A_- = [A_+, A_-]$.
This is cancelled by the $I$ term in eq. (\ref{virsym}),
since $D_\pm I = \pm A_\pm +
[A_\pm, I]$ has divergence $-[A_+, A_-]$.  Unfortunately, this does not seem to
generalize to curved space.

The equations $\partial^\mu (\rho A_\mu) = 0$ and $F_{+-} = 0$ can be solved to
give
\begin{eqnarray}
\partial_+ A_- &=& - {1\over 2} [A_+, A_-] - {1\over 2} \left({\partial_+
\rho\over\rho} A_- + {\partial_-\rho\over\rho} A_+\right)\nonumber \\
\partial_- A_+ &=& {1\over 2} [A_+, A_-] - {1\over 2}
\left({\partial_+\rho\over\rho} A_- + {\partial_- \rho\over\rho} A_+\right).
\end{eqnarray}
Using these relations, the divergence of $\delta' A_\mu$ contains three types
of terms proportional to $[A_+, A_-], A_+ \partial_- \rho$, and $A_- \partial_+
\rho$.  Let us examine the $[A_+, A_-]$ term.  We find
\begin{equation}
\partial_- (\rho \delta' A_+) + \partial_+ (\rho \delta' A_-) = {1 - t^2\over
t} {\tau \over 1 - \tau^2} F^2 [A_+, A_-]+ \dots,
\end{equation}
where the dots represent the $A_+ \partial_- \rho$ and $A_- \partial_+ \rho$
terms.  The coefficient of $[A_+, A_-]$ reduces to one in flat space, of
course, but it is a complicated function of $t$ and $x$ in curved space.  If
one attempts to add a variation $\delta^{(b)} g$ where such a factor multiplies
$I$, generalizing the second term in eq. (\ref{virsym}),
we can cancel the $[A_+, A_-]$
part of the divergence.  However, the $x$ dependence of this coefficient
results
in various additional pieces in $\partial^\mu (\rho \delta^{(b)} A_\mu)$.
Thus, cancelling one term generates new ones.  I have not succeeded in finding
a procedure to eliminate all of them.

\sectiono{Symmetric Space Models Coupled to Gravity}

The coupling of 2D SSM's to 2D gravity is given by formulas identical to those
for PCM's given in the preceding section.  In particular, the equations of
motion are given by eqs. (\ref{first}) -- (\ref{last}).
The only difference is that, just as for flat
space, the connection $A_\mu$ is given by $A_\mu = M^{-1} \partial_\mu M$ and
$M = g^{\dagger} g$ rather than $A_\mu = g^{-1} \partial_\mu g$.

\subsection{The symmetry transformations}

In view of our previous experience it is natural to conjecture that, just as in
flat space, the infinitesimal symmetry transformations that preserve the
equations $\partial^\mu (\rho A_\mu) = 0$ and $F_{+-} = 0$ for an SSM are given
by the same formulas as for PCM's.  These we found to be $g^{-1} \delta g = F
\eta$, where $F$ is given by eqs. (\ref{fform}) and (\ref{Fform})
and $\eta$ is given by eqs.
(\ref{etaeqn}) and (\ref{Xeqn}).  These result in the variation
\begin{equation}
\delta A_\mu = D_\mu (F\eta) + D_\mu (FM^{-1} \eta^{\dagger} M).
\label{dAgravSSM}
\end{equation}
The first term is identical to the PCM case and has already been shown to give
a vanishing contribution to $\partial^\mu (\rho \delta A_\mu)$.  Thus, only the
contribution of the second term needs to be checked.  Of course, we know from
Section 2.3 that it gives a vanishing contribution in the flat space ($\rho$
= constant) limit.  The generalization to curved space works because of the
identity
\begin{equation}
D_\pm (M^{-1} \eta^{\dagger} MF) = \pm {1 \over \rho}
\partial_{\pm}(\tau \rho FM^{-1} \eta^{\dagger} M).
\end{equation}
The derivation of this formula depends on the relation
\begin{equation}
\rho\partial_\pm F = \pm \partial_{\pm} (\tau\rho F).
\end{equation}
This formula differs from eq. (\ref{taudown})
in that $\tau$ appears now in the numerator
rather than the denominator.  Remarkably, both formulas are true because $\tau
\rho F$ and $\rho F/\tau$ differ by a constant.  Specifically,
\begin{equation}
\rho f \tau - {\rho f\over\tau} = - {4\over \kappa}.
\end{equation}
Thus, $g^{-1} \delta g = F\eta$ is indeed a classical symmetry of SSM's coupled
to gravity.

\subsection{The symmetry algebra}

The crucial formula required for commuting two symmetry transformations is
$\delta_1 X_2$.  As usual, it is found by solving
\begin{equation}
(\partial_\pm + \alpha_{2\pm} A_\pm)\delta_1 X_2 + \alpha_{2\pm}
(\delta_1 A_\pm) X_2 = 0,
\end{equation}
subject to the boundary condition that the variation vanishes
at $x = x_0$.  The result has the structure
\begin{equation}
(\delta_1 X_2)_{\rm SSM} = (\delta_1 X_2)_{\rm PCM}
+ (\delta_1 X_2)_{\rm extra},
\end{equation}
where $(\delta_1 X_2)_{\rm PCM}$ refers to eq. (\ref{dXPCM}),
the result found in Section
3.3.  This term is attributable to the term $D_\mu (F\eta)$ in
eq. (\ref{dAgravSSM}).
Thus, the rest of the answer requires solving the differential
equations with the source
$\delta_1 A_\pm = D_\pm (F_1 M^{-1} \eta_1^{\dagger} M)$.
The answer that results is
\begin{equation}
(\delta_1 X_2)_{\rm extra} =
{\tau_1\tau_2\over 1 - \tau_1\tau_2} F_1 M^{-1} \eta_1^{\dagger}
M X_2 + {t_1t_2\over t_1 t_2 - 1} X_2 M_0^{-1} \epsilon_1^{\dagger} M_0.
\end{equation}
The first term solves the inhomogeneous differential equation, and the second
one is a solution of the homogeneous equation chosen to ensure that the
variation
vanishes at the base point $x_0$.  In verifying this result, one finds the
formula by similar algebra to the flat space case.  However, an additional
identity
\begin{equation}
\partial_\pm \left({\tau_1\tau_2\over 1 - \tau_1\tau_2} F_1\right) +
\alpha_{2\pm} \partial_\pm F_1 = 0,
\end{equation}
which is trivial in flat space, also needs to be satisfied.  Even though it was
expected to work, it still seems remarkable that it is true.

Now we can compute the commutator of two symmetry transformations.  As in the
flat space case, we find that
\begin{equation}
[\delta_1, \delta_2] g = {t_1 \delta(\epsilon_{12}, t_1) - t_2 \delta
(\epsilon_{12}, t_2)\over t_1 - t_2} g + \delta' g + \delta^{\prime\prime} g.
\end{equation}
The term shown explicitly is the PCM result, unchanged from flat space.  The
$\delta^{\prime\prime} g$ term is also unchanged from flat space
\begin{equation}
\delta^{\prime\prime}g = {t_1 t_2\over 1 - t_1 t_2 }
\left( \delta(\epsilon'_{21},t_1) - \delta(\epsilon'_{12}, t_2) \right)g,
\end{equation}
The $\delta' g$ term is again a local $H$ transformation $(\delta' g = h_{12}
g.)$  However, the formula for $h_{12}$ is modified from the flat space one
given in eq. (\ref{htrans})
\begin{equation}
h_{12} = {\tau_1 \tau_2\over 1 - \tau_1 \tau_2} F_1 F_2
\Big[(g^\dagger)^{-1} \eta_1^\dagger M \eta_2 g^{-1}
+ g \eta_1 M^{-1} \eta_2^\dagger g^\dagger\Big] - {\rm h.c.}.
\end{equation}
As usual, it can be ignored.

Now modes $\Delta_n (\epsilon)g$ can be defined by the usual contour integral.
The integral has the same singularity structure as in the PCM case, and
therefore, the contours are taken to enclose the branch cuts as in that case.
Just as for PCM's, the algebra is unchanged from the flat space case
(aside from irrelevant local $H$ terms).
Thus, it remains the same as described in Section 2.3.

\sectiono{Conclusion}

In this paper we have shown that the affine Lie algebra symmetries of
classical PCM's and SSM's in flat
two-dimensional space-time remain symmetries in arbitrary
gravitational backgrounds. On the other hand, we were unable to
demonstrate that the Virasoro symmetries of the flat-space theories
survive in gravitational backgrounds. Perhaps it will become clearer
how this issue is resolved when the analysis is extended to the quantum
theory, and a Sugawara construction of the internal stress tensor that
generates the Virasoro symmetries can be attempted. However, since the
quantum theory is only expected to contain a discrete subgroup of duality
symmetries, which would not be accessible in terms of generators, it may be
necessary to understand finite group transformations first.
Another interesting direction to explore is the extension of our analysis
to the case of a circular spatial dimension. This is
an essential preliminary to the study of the hyperbolic symmetry
algebras that
are expected to appear after compactification of all spatial dimensions.

I am grateful to A. Sen for discussions and for reading the manuscript.

\end{document}